\documentclass[12pt,preprint]{aastex}

\shorttitle{Haze on Titan}

\shortauthors{Liang et al.}

\begin{document}

\title{Photolytically generated aerosols in the mesosphere and thermosphere of Titan}

\author{Mao-Chang Liang\altaffilmark{1,2}, Yuk L. Yung\altaffilmark{2}, and Donald E. Shemansky\altaffilmark{3}}

\altaffiltext{1}{Research Center for Environmental Changes, Academia
Sinica, Taipei 115, Taiwan}

\altaffiltext{2}{Division of Geological and Planetary Sciences,
California Institute of Technology, Pasadena, CA 91125, USA}

\altaffiltext{3}{Planetary and Space Science Division, Space
Environment Technologies, Pasadena, CA 91107, USA}

\begin{abstract}
Analysis of the Cassini Ultraviolet Imaging Spectrometer (UVIS)
stellar and solar occultations at Titan to date include 12 species:
N$_{2}$ (nitrogen), CH$_{4}$ (methane), C$_{2}$H$_{2}$ (acetylene),
C$_{2}$H$_{4}$ (ethylene), C$_{2}$H$_{6}$ (ethane), C$_{4}$H$_{2}$
(diacetylene), C$_{6}$H$_{6}$ (benzene), C$_{6}$N$_{2}$
(dicyanodiacetylene), C$_{2}$N$_{2}$ (cyanogen), HCN (hydrogen
cyanide), HC$_{3}$N (cyanoacetylene), and aerosols distinguished by
a structureless continuum extinction (absorption plus scattering) of
photons in the EUV. The introduction of aerosol particles, retaining
the same refractive index properties as tholin with radius $\sim$125
\AA\ and using Mie theory, provides a satisfactory fit to the
spectra. The derived vertical profile of aerosol density shows
distinct structure, implying a reactive generation process reaching
altitudes more than 1000 km above the surface. A photochemical model
presented here provides a reference basis for examining the chemical
and physical processes leading to the distinctive atmospheric
opacity at Titan. We find that dicyanodiacetylene is condensable at
$\sim$650 km, where the atmospheric temperature minimum is located.
This species is the simplest molecule identified to be condensable.
Observations are needed to confirm the existence and production
rates of dicyanodiacetylene.
\end{abstract}


\section{Introduction}
Titan is Nature's laboratory for organic synthesis. The major
molecules in the atmosphere are N$_2$ and CH$_4$. The coupled
chemistry between nitrogen and carbon leads to high abundances of
nitrogen/carbon compounds, such as hydrogen cyanide \citep[see,
e.g.,][]{Yung84,Coust03,Wilson04}. This is caused primarily by the
low gravity of Titan which allows hydrogen to escape readily
\citep[e.g.,][]{Yelle06}, resulting in low hydrogen abundance and
rich hydrocarbon production. When the order of hydrocarbons and
nitriles is large enough, they condense to form aerosols and
precipitate. The stratosphere is a major region for the production
of haze \citep[e.g.,][]{McKay01}. The estimated vertically
integrated rate is 0.5-2$\times$10$^{-14}$ g cm$^{-2}$ s$^{-1}$ with
haze formation taking place in the 300 km to 500 km region. Total
mass loading according to \citet{McKay01} was about 250 mg m$^{-2}$,
made by molecules with a C/N ratio of 2-4 and a C/H ratio of about
unity. The present work obtains about 100 mg m$^{-2}$ (assuming
density 3 g cm$^{-3}$) above 300 km.

\section{Cassini UVIS Observations}
On 13 December 2004, the Cassini UVIS recorded the occultation of
two stars, $\lambda$ Sco (Shaula; latitude $-$36$^\circ$) and
$\alpha$ Vir (Spica; latitude range of $+$63$^\circ$ and
$+$48$^\circ$), near the end of the second Titan flyby
(T$_{\mathrm{B}}$) \citep[details are referred
to][]{Shem05,Shem06,Sheme07}. The fully reduced results from
$\lambda$ Sco only are referenced here. The vertical distribution of
the aerosol component as the terminal product of N$_{2}$/CH$_{4}$
physical and chemical processes, is of primary interest to this
Letter. The spectral region 1850-1900 {\AA} (SP1) is effectively
free of the measurable hydrocarbon and cyano species in this
atmosphere apart from aerosols, and this region is used
photometrically to trace the aerosol structure. The vertical profile
of extinction for SP1 below 1000 km was carried out at the highest
possible ray-height resolution (3-5 km) in order to reveal possible
structure in the distribution.

\begin{figure*}
 \epsscale{1} \plotone{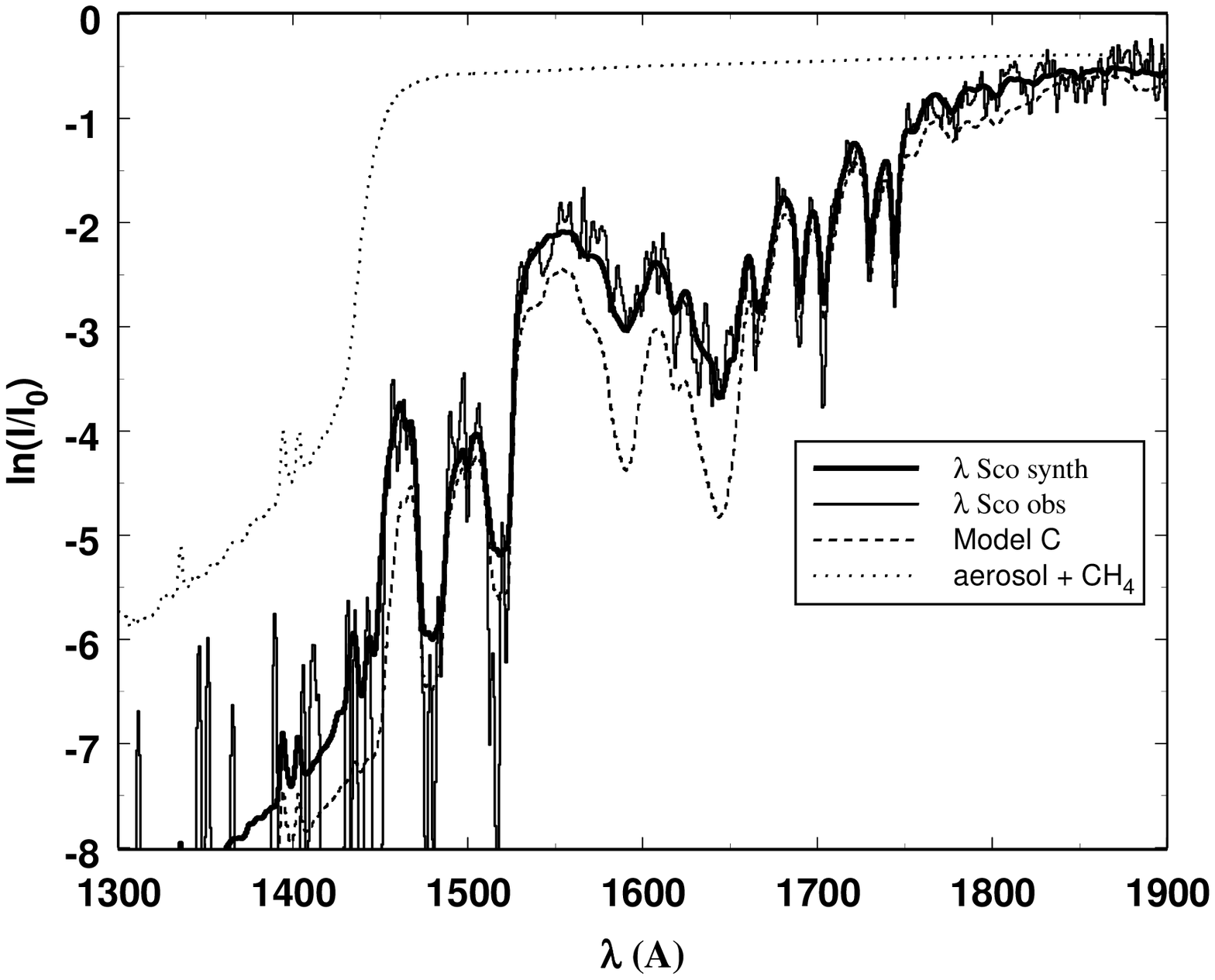}
\vspace*{-2.5in}\caption[]{The transmission spectrum of the UVIS
$\lambda$ Sco occultation integrated over the impact parameter 514
km to 537 km (light line), compared to the best-fit synthesis
\citep{Sheme07} using the combined identified species (heavy line),
and to model C from the present physical chemistry code (dashed
line). The dotted line shows the aerosol component combined with the
CH$_{4}$ absorber in this reduction. The CH$_{4}$ absorber
($\lambda$ $<$ 1490 {\AA}) is included with aerosol here as a means
of including a large part of the impact of instrument point spread
function on the fitting process.  The optical depth in region SP1 is
entirely attributed to aerosol extinction; The small difference
between the observed data and the aerosol component at SP1 is an
artifact of the UVIS EUV instrument point spread function. The
abundances of the species for this case are given in Table
\ref{model_summary}. See text. \label{vtrans}}
\end{figure*}

The best fit to the transmission spectrum at impact parameter h $=$
514 km is shown in Figure \ref{vtrans} as an example of spectral
reduction. Table \ref{model_summary} shows the extracted
line-of-sight abundances from this element of the occultation.
Figure \ref{tholdens} shows the aerosol density vertical
distribution (heavy dots) derived from the reduction of the
$\lambda$ Sco occultation, compared to the derived CH$_{4}$ profile
(dashed line). The reduced data extend from h $=$ 330 km to 970 km
where signal noise terminates the reduction (see Shemansky et al.
2007 for details). The remarkable property of the aerosol
distribution is the sudden departure from tracking the CH$_{4}$
abundance at h $=$ 468 km toward higher altitudes. The interval
between 468 km and 550 km where abundance remains approximately
constant, is interpreted as a primary aerosol source region. The
density extraction is a direct deconvolution of the abundance
distribution. The derivation is based on assumed spherical
uniformity and assumed uniformity in composition with altitude. The
observations show measurable simple scattering in the 1800 {\AA}
region at h $=$ 1250 km, indicating that aerosols are extensively
distributed into the thermosphere.

The cross sections for extinction of UV light by aerosols were
computed using the complex refractive index measured by
\citet{Khare84} for solid state tholins, and the scattering code of
\citet{Mish98}. Representative complex indices of refraction at 588,
1215, 1631 and 2384 {\AA} are (0.963, 0.62), (1.74, 0.37), (1.65,
0.24) and (1.68, 0.21), respectively. Assuming a mean radius of 125
{\AA} for aerosols, the corresponding extinction cross sections are
2.4$\times$10$^{-12}$, 3.9$\times$10$^{-13}$, 2.0$\times$10$^{-13}$
and 1.2$\times$10$^{-13}$ cm$^{2}$, respectively. The actual
computation used in this work has a finer wavelength grid. Note that
the current UVIS data set no constraint on the vertical variation of
the optical property of aerosols.

Extinction of the EUV stellar photons is dominated by CH$_{4}$ in
the 1100 {\AA} to 1400 {\AA} region. At longer wavelengths the
structure is a combination of the higher order hydrocarbon and cyano
species. The C$_6$H$_6$ cross section peaks at 1759.9-1815.1 \AA,
blended primarily with C$_2$H$_4$. Dicyanodiacetylene has a cross
section peak in the SP1 spectral region. Dicyanodiacetylene and
benzene have not been detected in the absorption spectra
\citep{Sheme07}. Aerosol extinction is detectable at $\sim$970 km in
the transmission spectra and dominates all absorbers at all
wavelengths in the UVIS except for CH$_4$, at altitudes below
400-450 km.

Figure \ref{vtrans} shows the contribution of aerosol extinction to
the total measured optical depth at 514 km. In the spectral region
SP1 aerosol extinction is entirely responsible for the optical
depth. The measurable spectral region for extracting the aerosol
component is 1500-1900 {\AA}, where the wavelength dependence of
extinction shows a proportionality to $\lambda^{-1.5}$. The Voyager
and Cassini photometric observations in the UV spectral region
\citep{Porco05,West06} have revealed the presence of detached haze
layers at Titan. The Cassini results show the presence of a
latitudinally uniform detached layer near 500 km in forward
scattered 3380 {\AA} photons. The relationship of this phenomenon to
the aerosols identified here requires further investigation. A
comparison to Voyager results \citep{Smith82} shows that the major
differences are that the apparent strong extinction by aerosols
takes effect about 100 km higher for Voyager and significantly more
extinction is evident for Voyager in the 700-1000 km region. The
Voyager data show a broad extinction maximum near 770 km.

\section{Photochemical Modeling}

\begin{figure*}
\epsscale{1} \plotone{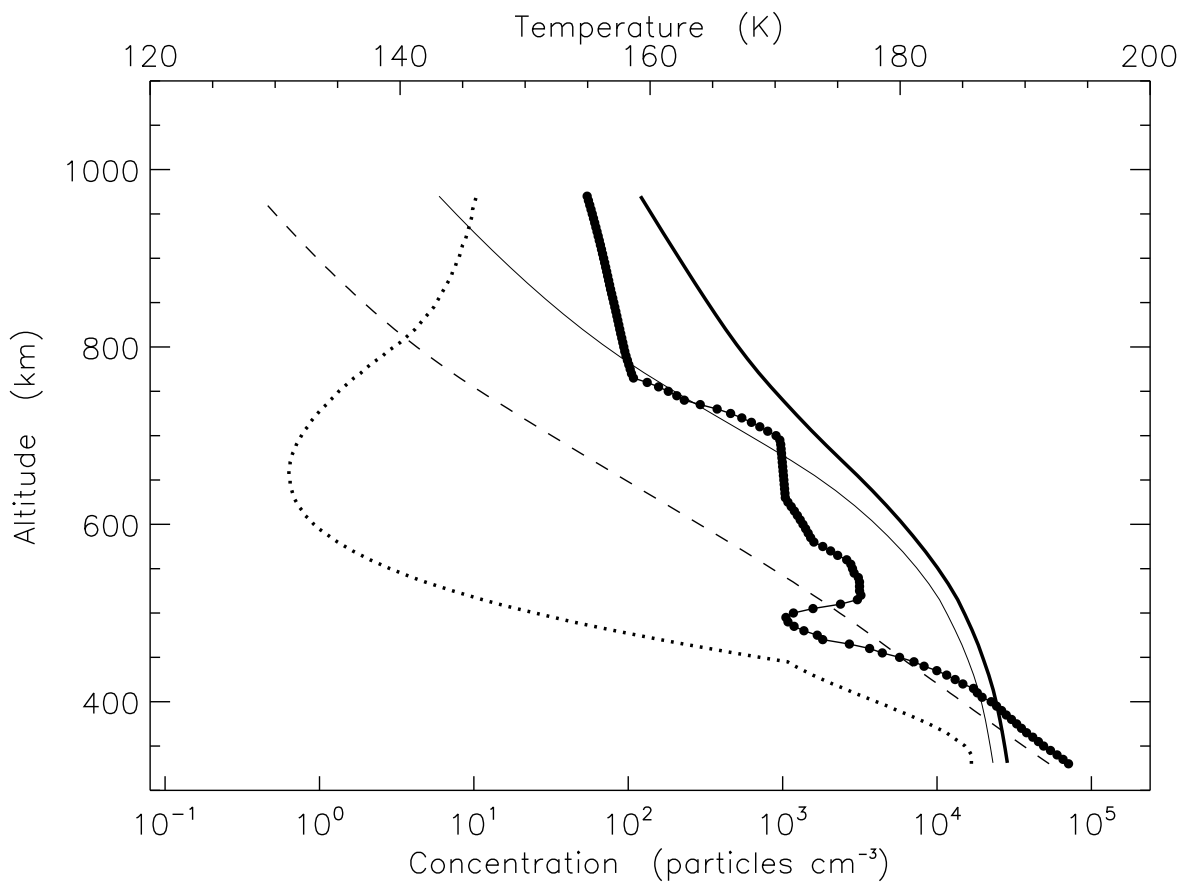} \caption[]{Aerosol density (heavy
dots) derived from the UVIS $\lambda$ Sco occultation compared to
the CH$_{4}$ (dashed line) scaled by 10$^{-9}$
\citep{Shem05,Shem06}. The increase of the mixing ratio of the UVIS
aerosols through the mesosphere to at least 1000 km implies that the
production of aerosols must take place at significant rates
throughout the mesosphere and thermosphere. The UVIS derived
temperature profile is shown by the dotted line, which reflects a
correction to the one presented by \citet{Shem05}. The model aerosol
profiles are shown by the thin and solid lines (see text). The over-
and underestimations are due to the fact that we assume a constant
sedimentation velocity of 0.25 cm s$^{-1}$, calculated at 535 km
\citep{Cabane1992}. \label{tholdens}}
\end{figure*}

Vertical profiles of the major species have been calculated using a
photochemical model. The photochemical reactions are taken from
\citet{Yung84}, \citet{Yung87}, and \citet{Moses00}. The chemical
scheme to C$_6$N$_2$ is hypothesized to be similar to that to
C$_4$N$_2$, as derived by \citet{Yung87} and summarized in Table
\ref{reactions}. The temperature profile is based on the Cassini
measurements (Figure \ref{tholdens}, dotted line). The vertical eddy
mixing profile is taken from \citet{Yung84}. The model simulation is
diurnally averaged at low latitude. The incident UV flux is the mean
between solar maximum and minimum. Table \ref{model_summary}
provides a summary of model results. Sensitivity to the selection of
hydrocarbon kinetics and that of kinetics and vertical eddy
coefficients are shown by models D and WA04, respectively.

We fix the N$_2$ abundance to that derived from the Cassini
measurements. The model starts with a hydrostatic atmosphere. With
the prescribed vertical diffusion coefficients and taking the
photolysis of CH$_4$ into account, the abundance of CH$_4$ is
overestimated (see Table \ref{model_summary}), compared with the
measurements. To bring the model into better agreement with the
observations, we introduce an {\it ad hoc} advection which
transports species other than N$_2$ downward. The wind is prescribed
with strength proportional to the inverse of the square root of
atmospheric density. The wind speed reaches -20 cm s$^{-1}$ at the
top of the model atmosphere ($\sim$1500 km). The assumed downward
wind is qualitatively consistent with global circulation that has a
downward transport at mid to high latitudes
\citep[e.g.,][]{Lebonn01}. A comparison with modeled CH$_4$
abundances between models A and C is shown in Figure \ref{profiles}.
We note that dynamics plays an important role in distributing
photochemical products \citep[e.g.,][]{Lebonn01}, especially in the
regions above $\sim$500 km (the regions of interest to this work)
where the transport time is, in general, shorter than the chemical
removal time of hydrocarbon and cyano species
\citep[e.g.,][]{Wilson04}; current simulations coupled with dynamics
and photochemistry are limited to the regions below $\sim$400 km
\citep{Lebonn01}. The latitudinal variations of hydrocarbon and
cyano abundances \citep[e.g.,][]{Flasar05} are the consequence of
atmospheric transport and photochemical processes.

The modeled profiles of HCN, HC$_3$N, C$_6$H$_6$, and C$_6$N$_2$ are
presented in Figure \ref{profiles}. Calculated abundances are
compared to extraction from observation at 514 km in Table
\ref{model_summary}. Comparisons between models and observations at
other impact parameters will be deferred to a later paper. The
results for five variations on the model at this impact parameter
are given in Table \ref{model_summary}. In general, our base models
(models A and B) overestimate the abundances of hydrocarbons by as
much as 10 times. An indication of the difference in the predicted
model C and observed optical depth spectra is shown in Figure
\ref{vtrans}. Model C (Table \ref{model_summary}) is too high
relative to measured abundance in C$_{2}$H$_{6}$ and C$_{4}$H$_{2}$.
The modeled C$_{6}$N$_{2}$, C$_{6}$H$_{6}$, and HC$_{3}$N are also
well above the upper limits set by observation. There are two ways
of reducing the abundances. (1) Faster transport of photochemical
products to the lower atmosphere as in the model of \citet{Wilson04}
(model WA04). Comparing the transport time constant with the
chemical destruction time, the abundances of C$_2$H$_2$, C$_2$H$_6$,
HCN, and C$_2$N$_2$ have sensitivity to transport and those of
C$_2$H$_4$, C$_4$H$_2$, C$_6$N$_2$, C$_6$H$_6$, and HC$_3$N at
$\sim$500 km are close to being in photochemical equilibrium
\citep[e.g.,][]{Wilson04}. (2) Relatively rapid two-body
physicochemical processes forming aerosols as simulated in models C
and D. The loss rates in these models are assumed to be proportional
to the physical collision rates between aerosols (with radius 125
\AA) and molecules; adsorption reactions are assumed for all
photochemical species listed in Table \ref{model_summary}. The
aerosol density is from Figure \ref{tholdens}. The adsorption
efficiency for this absolute loss, an assumed value of 0.01, yields
the concentrations shown in Table \ref{model_summary}. As described
below, this process is required for explaining the aerosol abundance
shown in Figure \ref{tholdens}.

\section{Discussion}

The source of aerosols has long been a puzzle in the atmosphere of
Titan. It is generally believed that the synthesis of increasingly
complex hydrocarbon and nitrogen compounds will eventually lead to
saturation, resulting in coagulation and precipitation. However, the
chemical composition of the condensible species has not yet been
established. In this Letter, we propose that the simplest
condensible compound is C$_6$N$_2$. The abundances of the higher
order species in the UVIS observations are significantly lower than
the present model calculations. Assuming that the model conversion
rates for the CH$_{4}$ source are basically correct, there is an
implied loss rate for these species that is substantially higher
than the model provides. The model calculation contains an absolute
loss to the measured aerosols using a conservatively small
adsorption probability. If a significant fraction of the implied
loss is delivered to the production of aerosols, the model can be
adjusted by assuming a larger irreversible adsorption probability
and a consequent higher precipitation rate for the aerosols. This
will bring the model abundances into better conformance with
observation, but will not necessarily resolve differences in
partitioning, and may not resolve issues raised in regard to rate
limits for the mass flow process that cycles to the surface. We
propose that the aerosol formation is initiated by the condensation
of C$_6$N$_2$ and adsorption to external meteoritic dust; both serve
seeds for aerosol formation. The subsequent physical processes of
adsorption on the existing aerosols and (photo)chemistry converting
these clusters into refractory tholins constitute the production and
maintenance of the aerosol distribution. \citet{Hunten} has recently
proposed that haze is a major sink of ethane at Titan. The process
of formation of aerosols requires the stable adsorption of the
higher order hydrocarbon and nitrile species.

\begin{figure*}
 \epsscale{1} \plotone{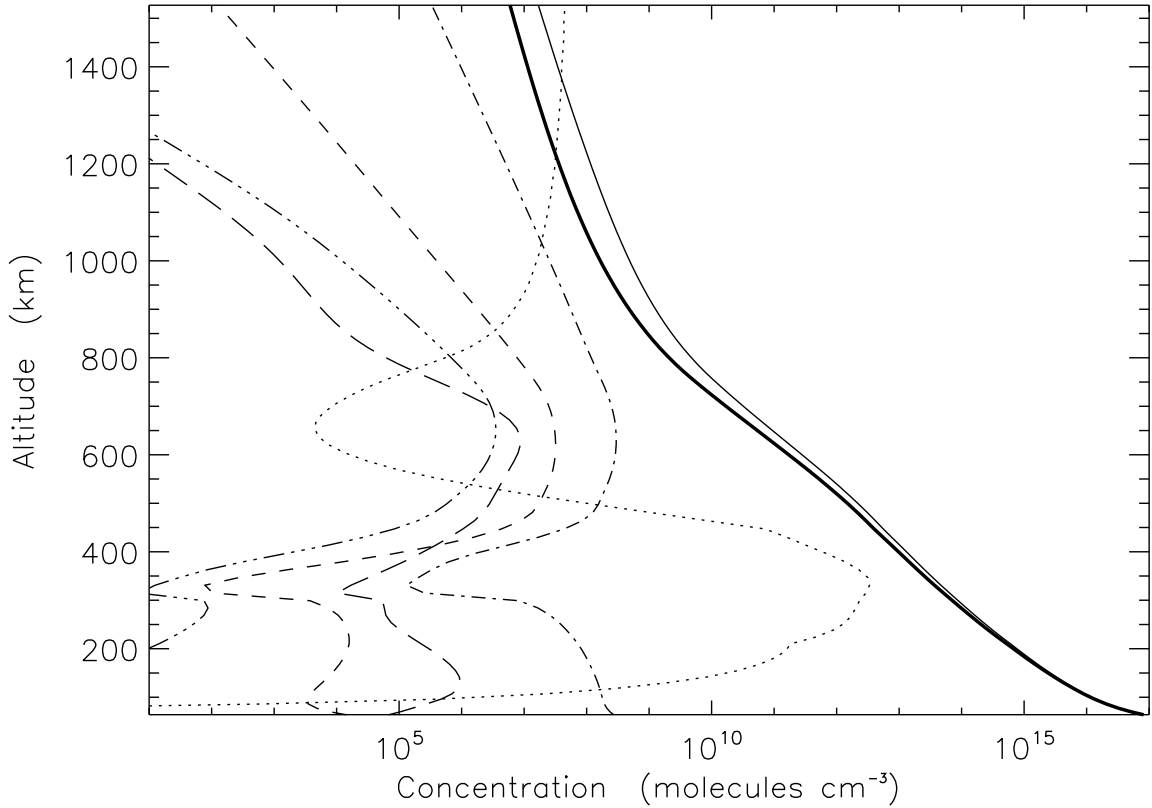}
\caption[]{Modeled (model C) vertical profiles for CH$_4$ (thick
solid), HC$_3$N (dashed), HCN (dash-dotted), C$_6$N$_2$
(triple-dot-dashed), and C$_6$H$_6$ (long-dashed). Thin solid line
represents modeled CH$_4$ by model A. The saturation density of
C$_6$N$_2$ extrapolated from high temperature measurements
\citep[295-369 K,][]{Saggiomo57} is shown by dotted line. The
resulting H$_2$ (3$\times$10$^{-3}$) and CH$_4$ (2.3\%) mixing
ratios at 1174 km and H$_2$ escape flux (7$\times$10$^{9}$ molecules
cm$^{-2}$ s$^{-1}$) at the top are in good agreement with the
observations (4$\pm$1$\times$10$^{-3}$, 2.7$\pm$0.1\%, and
1.2$\pm$0.2$\times$10$^{10}$ molecules cm$^{-2}$ s$^{-1}$,
respectively) \citep{Yelle06}. \label{profiles}}
\end{figure*}

The production rate of C$_{6}$N$_{2}$ in the model limits the
production rate of aerosols from this direct path. The saturation
density of dicyanodiacetylene shown in Figure \ref{profiles} shows
that condensation can take place between $\sim$550 and 800 km.  The
volume production rate of C$_6$N$_2$ in this region is quite uniform
($\sim$1 molecules cm$^{-3}$ s$^{-1}$); the column integrated
(550-800 km) rate is $\sim$10$^7$ molecules cm$^{-2}$ s$^{-1}$, or
$\sim$2$\times$10$^{-15}$ g cm$^{-2}$ s$^{-1}$. This contribution is
a small fraction of the total aerosol production, but is extremely
important as a source of condensation nuclei. The maximum rate of
aerosol production, however, is set by the total photolysis rate of
CH$_4$ which is on the order of 10$^{10}$ molecules cm$^{-2}$
s$^{-1}$, or 2$\times$10$^{-13}$ g cm$^{-2}$ s$^{-1}$ of carbon, a
result that is independently corroborated by the H$_2$ escape flux
\citep{Yelle06}. The rate of removal of molecules by the existing
particles in model C is close to this maximum rate. However, the
resulting aerosol profile (thin solid line in Figure 2)
underestimates the observed aerosol abundances above $\sim$800 km,
suggesting an additional source at the top of the atmosphere,
equivalent to a downward flux of 5$\times$10$^{-14}$ g cm$^{-2}$
s$^{-1}$ (thick solid line). This flux is consistent with that
inferred from the Cassini Ion Neutral Mass Spectrometer measurements
\citep{Waite2007} on the basis of ion chemistry not considered here.
Note that a sedimentation velocity of 0.25 cm s$^{-1}$ has to be
imposed in order to match the observed aerosol profile; this implies
that the radius of the aerosols must be $\sim$125 \AA\ in the
mesosphere and thermosphere.

We emphasize that there are significant uncertainties in rate
process quantities and stronger constraints are needed. Laboratory
measurements for the adsorption rates on aerosols in collision with
the high order molecules are required to verify this process, and to
provide a better constraint to the aerosol mass loading in the
atmosphere of Titan. In addition, the photochemical paths to high
order hydrocarbon and nitrile compounds such as C$_6$N$_2$ and
C$_6$H$_6$ are speculative. Atmospheric dynamics in the mesosphere
of Titan also plays a central role in molecule/aerosol mixing.
Further laboratory measurements and Cassini observations will
provide valuable information in refining our understanding of the
chemical, dynamical, and microphysical processes in the atmosphere
of Titan.

\acknowledgements This research was supported by NASA grant
NNG06GF33G and Cassini grant JPL.1256000 to the California Institute
of Technology. DES acknowledged support by NASA grant NNG06GH76G and
Cassini UVIS Program contract 1531660 to Space Environment
Technologies.

\clearpage

\clearpage

\clearpage

\clearpage

\begin{deluxetable}{lrccccc}
\tablecolumns{5} \tablecaption{Summary of Model Results
\label{model_summary}} \tablewidth{0pt} \tablehead{
\multicolumn{1}{c}{Molecule} & \multicolumn{1}{c}{Cassini} &
\multicolumn{1}{c}{Model A} &\multicolumn{1}{c}{Model B} &
\multicolumn{1}{c}{Model C} &\multicolumn{1}{c}{Model D}
&\multicolumn{1}{c}{WA04} }
\startdata
 N$_{2}$ ($\times$10$^{21}$)    &   5.8 &   5.8 &   5.8 &  5.8 &  5.8 &  5.8 \\
 CH$_4$ ($\times$10$^{19}$)     &   6.0 &   15  &   9.4 &  9.5 &  9.7 &  13   \\
 C$_2$H$_2$ ($\times$10$^{17}$) &   2.1 &   15  &   9.1 &  1.9 &  1.7 &  1.5  \\
 C$_2$H$_4$ ($\times$10$^{16}$) &   4.0 &   9.3 &   5.7 &  4.0 &  2.0 &  3.4  \\
 C$_2$H$_6$ ($\times$10$^{16}$) &   7.0 &   200 &   110 &  17  &  9.2 &  20   \\
 HCN ($\times$10$^{17}$)        &   1.0 &   5.6 &   3.7 & 0.69 & 0.53 & 0.017 \\
 C$_4$H$_2$ ($\times$10$^{15}$) &   4.5 &   59  &   37  &   12 &  2.1 &  41   \\
 C$_6$N$_2$ ($\times$10$^{14}$) &$<$1.0 &   15  &   16  &  5.7 &  8.0 &  \nodata   \\
 C$_6$H$_6$ ($\times$10$^{14}$) &$<$1.4 &   18  &   13  &  12  & 0.041&  0.21 \\
 HC$_3$N ($\times$10$^{15}$)    &$<$3.9 &   27  &   25  &  6.9 &  7.8 &  0.96  \\
 C$_2$N$_2$ ($\times$10$^{15}$) &$<$4.0 &  0.84 &   1.2 & 0.34 & 0.42 &  0.00035  \\
 Tholin ($\times$10$^{11}$)     &   4.6 & \nodata & \nodata & \nodata & \nodata & \nodata \\
\enddata
\tablecomments{Values are line-of-sight column integrated
abundances, in molecules cm$^{-2}$, reported by matching the
observed N$_2$ abundance. Model A: hydrostatic atmosphere. Model B:
non-hydrostatic atmosphere, an {\it ad hoc} downward wind and
extinction due to the derived tholins are assumed (see text). Model
C: same as Model B but also with additional sinks for the tabulated
nine photochemical species (see text). Model D: same as model C but
with the updated hydrocarbon chemistry from \citet{Moses05}. WA04:
model results from \citet{Wilson04}. Note that in this Letter, the
microphysical processes of C$_6$N$_2$ are not solved
self-consistently; and hence, the tabulated abundances of C$_6$N$_2$
do not reflect the removal by condensation.}

\end{deluxetable}

\clearpage

\begin{deluxetable}{llllll}
\tablecolumns{6} \tablecaption{Chemical Reactions to C$_6$N$_2$
\label{reactions}} \tablewidth{0pt} \tablehead{
\multicolumn{1}{c}{Label} & \multicolumn{1}{c}{Reactants} &
\multicolumn{1}{c}{} & \multicolumn{1}{c}{Products}
&\multicolumn{1}{c}{} & \multicolumn{1}{c}{Rate
Coefficients\tablenotemark{a}}}
\startdata
R454 & HC$_5$N + h$\nu$           & $\rightarrow$ & C$_4$H + CN                &  & = J(HC$_3$N + h$\nu$ $\rightarrow$ C$_2$H + CN); (1) \\
R455 & HC$_5$N + h$\nu$           & $\rightarrow$ & C$_5$N + H                 &  & = J(HC$_3$N + h$\nu$ $\rightarrow$ C$_3$N + H); (1) \\
R456 & C$_6$N$_2$ + h$\nu$        & $\rightarrow$ & C$_5$N + CN                &  & = J(C$_4$N$_2$ + h$\nu$ $\rightarrow$ C$_3$N + CN); (2) \\
R492 & C$_3$N + HC$_3$N           & $\rightarrow$ & C$_6$N$_2$ + H             &  & = k(C$_2$H + C$_2$H$_2$ $\rightarrow$ C$_4$H$_2$ + H); (3) \\
R495 & CN + C$_4$H$_2$            & $\rightarrow$ & HC$_5$N + H                &  & = k(CN + C$_2$H$_2$ $\rightarrow$ HC$_3$N + H); (2) \\
R496 & CN + HC$_5$N               & $\rightarrow$ & C$_6$N$_2$ + H             &  & = k(CN + HC$_3$N $\rightarrow$ C$_4$N$_2$ + H); (2) \\
R497 & C$_5$N + CH$_4$            & $\rightarrow$ & HC$_5$N + CH$_3$           &  & = k(C$_3$N + CH$_4$ $\rightarrow$ HC$_3$N + CH$_3$); (2) \\
R498 & C$_5$N + C$_2$H$_6$        & $\rightarrow$ & HC$_5$N + C$_2$H$_5$       &  & = k(C$_3$N + C$_2$H$_6$ $\rightarrow$ HC$_3$N + C$_2$H$_5$); (2) \\
\enddata

\tablenotetext{a}{Estimated from the quoted reactions. Units are
s$^{-1}$ for photolysis reactions (J) and cm$^3$~s$^{-1}$ for
two-body reactions (k). The photolysis rate coefficients are given
at the top of the model atmosphere. References: (1) \citet{Lebonn01}
and \citet{Wilson04}; (2) \citet{Yung87}; (2) \citet{Lebonn01} and
\citet{Opansky96}}
\end{deluxetable}

\end{document}